\documentclass{article}

\usepackage{subcaption}
\usepackage{float}
\usepackage{graphicx}
\usepackage{url}

\usepackage{authblk}

\begin{document}

\title{Towards Grassroots Peering at the Edge\footnote{This paper has been accepted at ACM M4IoT 2021. This document is the authors' copy which is -- aside from formatting -- identical to the one found in the ACM Digital Library.}}

\author[1]{David Bermbach}

\author[2]{Sergio Lucia}

\author[3]{Vlado Handziski}

\author[4]{Adam Wolisz}

\affil[1]{TU Berlin \& ECDF\\Mobile Cloud Computing Research Group\\\texttt{db@mcc.tu-berlin.de}}

\affil[2]{TU Dortmund \& ECDF\\Process Automation Systems Group\\\texttt{sergio.lucia@tu-dortmund.de}}

\affil[3]{Reliable Realtime Radio\\\texttt{vlado.handziski@r3.group}}

\affil[4]{TU Berlin \& ECDF\\Telecommunication Networks Group\\\texttt{wolisz@tkn.tu-berlin.de}}

\date{}

\maketitle

\abstract{
Fog Computing allows applications to address their latency and privacy requirements while coping with bandwidth limitations of Internet service providers (ISPs).
Existing research on fog systems has so far mostly taken a very high-level view on the actual fog infrastructure.
In this position paper, we identify and discuss the problem of having multiple ISPs in edge-to-edge communication.
As a possible solution we propose that edge operators create direct edge-to-edge links in a grassroots fashion and discuss different implementation options.
Based on this, we highlight some important open research challenges that result from this.
}


\section{Introduction}
Today's applications are usually cloud-based due to the obvious advantages of the cloud -- elastic scalability, ease-of-use, (the illusion of) infinite resources, a pay-as-you-go model~\cite{mell2011nist,paper_bermbach_cloud_engineering}.
In practice, however, the cloud is often quite far away from end users or IoT devices, thus, resulting in latency problems for many applications~\cite{paper_bermbach_fog_vision}.
Furthermore, the volume of data produced by end users and devices already today far exceeds the bandwidth for transmitting this data to the cloud for processing~\cite{paper_zhang_gdp}.
Finally, storing and managing independent data sets in centralized locations (as in the cloud) may lead to misuse and undesired correlation of such data sets -- either accidentally or on purpose~\cite{paper_pallas_fog4privacy,grambow_public_2018}.
As a solution to these three challenges, the concept of fog computing has been proposed~\cite{bonomi2012fog}:
Existing resources in the cloud are combined with smaller edge nodes near the end users (on the ``edge'' of the network) which offer limited resources with low latency and high bandwidth access in a decentralized way (thus avoiding centralized data hubs~\cite{paper_pallas_fog4privacy}).
Beyond this, these edge resources may even be augmented with additional resources, e.g., small- to medium-sized data centers within the provider's core network, on the path from edge to cloud.
This all allows developers to combine all benefits and advantages of cloud and edge computing.
While there are multiple competing definitions in this area, the general idea of using multiple compute locations on the path from end users to the cloud is shared by all.
In this paper, we will refer to resources near the edge as edge nodes or edge computing, to cloud resources as cloud nodes, to compute resources on the path from edge to cloud as intermediary nodes, and to the combination of all as fog computing~\cite{paper_bermbach_fog_vision}.

In the last few years, much research has been done on fog systems, e.g.,~\cite{paper_hasenburg_broadcast_groups,confais2017object,paper_pfandzelter_zero2fog,happ2021joi,hasenburg_fogexplorer_2018,hasenburg_supporting_2018,paper_george_nanolambda,paper_pfandzelter_tinyfaas,paper_hasenburg_towards_fbase,techreport_hasenburg_2019,leidall2019edge}.
What they all have in common is that they assume a single provider of edge resources per location.
In practice, however, countries usually have multiple network carriers and third party entities that will provide independent edge infrastructure in parallel.
Particularly, this means that adjacent edge nodes will often be connected via the Internet backbone only.
To our knowledge, we are the first to focus on this problem.
In this position paper, we hence make the following contributions:
\begin{itemize}
	\item We identify and discuss the edge and fog computing implications of having several separate provider networks in parallel. We also discuss the implications of this on IoT applications (Section~\ref{sec:problem}).
	\item We propose an approach through which end users can partially address this problem in a grassroots fashion (Section~\ref{sec:grassroots}).
	\item We discuss open research directions that result from this (Section~\ref{sec:open}).
\end{itemize}

\section{The Problem of Separate Provider Networks\label{sec:problem}}
In this section, we will start by giving an overview of the usual, abstracted perspective on fog computing and discuss how it differs from actual fog deployments in practice.
Then, we will describe an IoT use case and discuss the implications of the difference between abstract and actual fog environments.

\subsection{The Contrast between Idealized and Actual Deployments in Fog Computing}

\begin{figure}
    \centering
    \begin{subfigure}{0.49\columnwidth}
        \centering
        \includegraphics[width=\columnwidth]{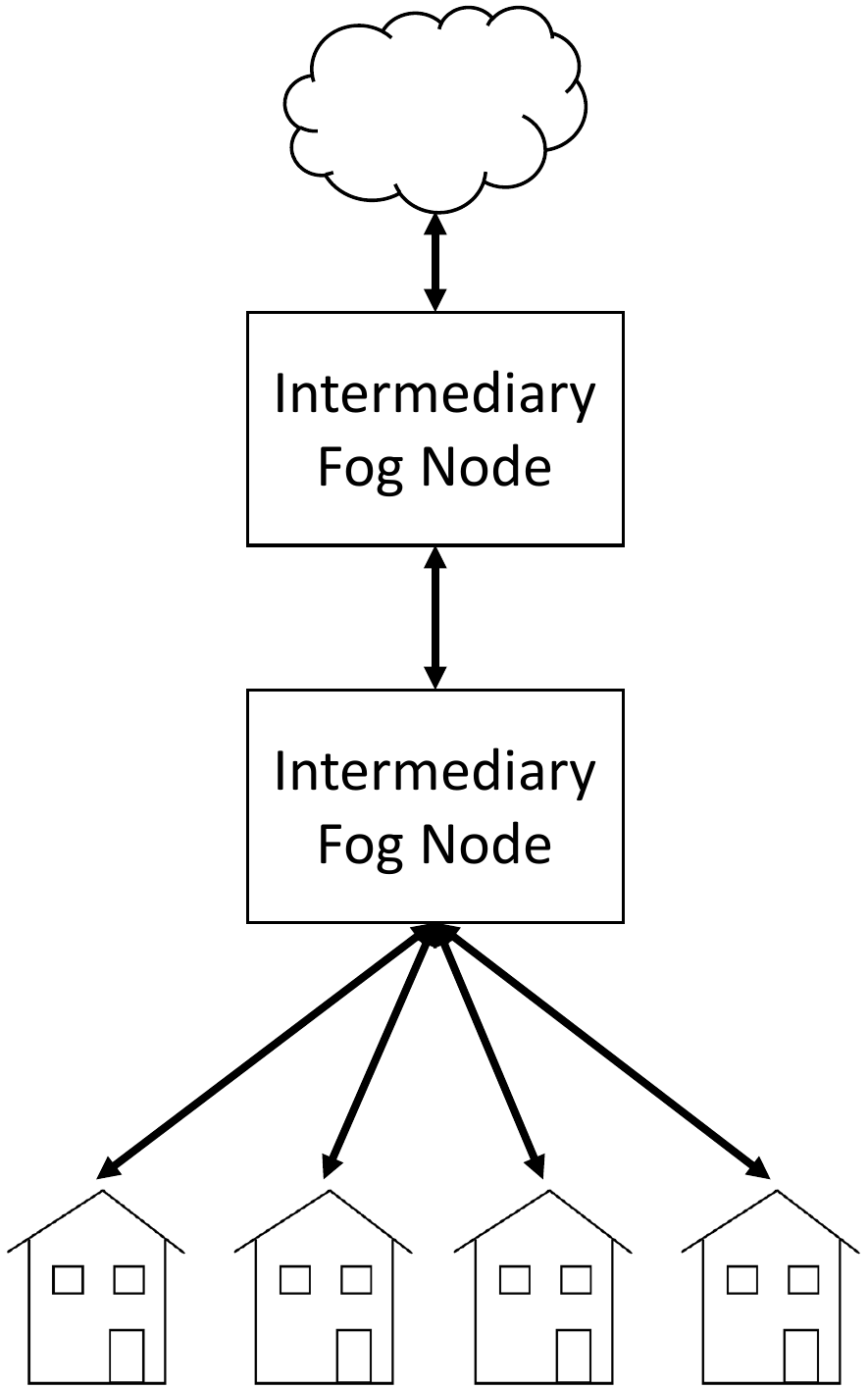}
        \caption{Implicitly assumed deployments in fog computing.}
        \label{fig:overview-assumed}
    \end{subfigure}
    \hfill
    \begin{subfigure}{0.4\columnwidth}
        \centering
        \includegraphics[width=\columnwidth]{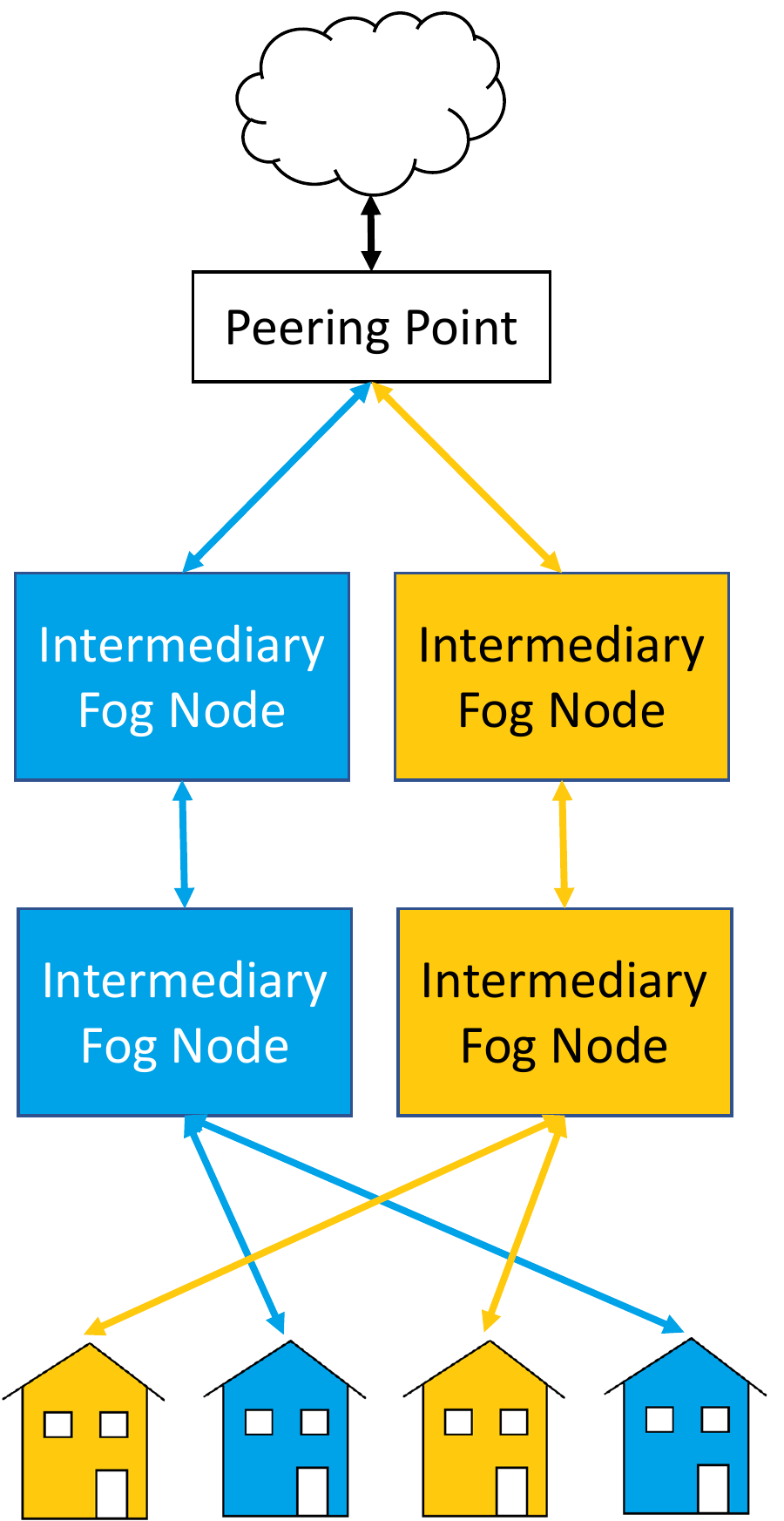}
        \caption{Actual deployments in fog computing.}
        \label{fig:overview-actual}
    \end{subfigure}
        \caption{In contrast to typical assumptions, actual fog deployments will in practice have multiple independent providers operate in parallel.}
    \label{fig:overview}
\end{figure}

As already discussed in our introduction section, the general high-level perspective on fog computing is that edge and cloud resources are combined with any resources on the path from edge to cloud.
Usually, this is visualized similar to Figure~\ref{fig:overview-assumed}: The fog forms a tree-like structure with the cloud as its root node, branches out through one or more layers of intermediary nodes, and has a high number of edge nodes as leaves, e.g.,~\cite{paper_bermbach_fog_vision,dustdar2019edge,paper_pfandzelter_functions_streams}.
Real fog deployments, however, could differ from this simplified model in a number of ways.

First, the intermediary nodes and the surrounding networks will in practice often be owned and operated by an Internet service provider (ISP) -- the ISP's core network; edge nodes may also belong to that ISP or a third party, e.g., on-premises compute resources.
While this provider core network is likely to resemble a tree near the edge, the upper layers of the ``tree'' will often be interconnected in various ways.
Usually, several different ISPs will be active in one geographic region, i.e., a country will usually have several independent core networks including the respective intermediary nodes.

Second, the cloud may not be the root of the tree as indicated, e.g., in Figure~\ref{fig:overview-assumed}, and is, in fact, usually owned and operated by another entity, e.g., one of the major cloud providers such as Amazon, Google, or Microsoft.
In fact, the cloud is not a single entity but rather a collection of data centers distributed worldwide.
We can, however, assume that usually only the logically nearest cloud data center will matter from an end user perspective and, thus, exclude the others from visualizations.
The key difference is that both the provider network and cloud data centers will usually form so-called autonomous systems which are connected via one or more peering points either directly or via the Internet backbone.

Third, the idealized fog usually assumes that edge-to-edge communication is fast and local.
In practice, most locations will have several independent ISPs that offer Internet access.
As a consequence, edge nodes -- no matter if owned/operated by the end user or the ISP -- will belong to the core network of the respective provider.
This situation is even more muddled when considering that ISPs often rent network capacity (including the last mile) from their competitors.
Overall, this means that physical neighbors and their respective edge nodes may belong to different ISP core networks.
In that case, their edge-to-edge communication will have to pass through the peering point(s) that their respective ISPs have chosen.
See Figure~\ref{fig:overview-actual} for a still high-level but more realistic overview of typical fog deployments.

Finally, cloud vendors also offer edge services, e.g., AWS Greengrass.
In that case, either the cloud vendor ships ready-to-use devices including hardware and software or provides the necessary software to let the customer run it on-premises.
In either case, the cloud vendor will operate the edge infrastructure which is then tightly coupled to the corresponding cloud offerings, making fog applications spanning different cloud vendors challenging.

\subsection{Use Case: Local Neighborhood IoT Sensor Data Sharing}
As a use case, consider a scenario of a ``smart'' neighborhood where individual neighbors want to share their local IoT sensor data with some or all of their neighbors.
One example would be temperature and precipitation sensors where individual measurements in a small geographic area will not differ significantly.
In such a setup, neighbors could save on installation and maintenance cost by sharing their data.
A second example would be the data from multiple anemometers or phonometers which could be used by citizen initiatives or researchers to collaboratively map out noise levels from a nearby highway or to study effects of infrastructure on wind speed.
Another class of applications are privacy-sensitive scenarios, often involving large data volumes from cameras or microphones.
For instance, someone visiting their neighbors might want to stream the data from an infant monitoring system to their neighbor's TV but might either not be comfortable with routing such data through the Internet backbone or might need more bandwidth (e.g., for HD resolution from multiple cameras) than available to them through their ISP's upstream bandwidth.

\begin{figure}
    \centering
    \begin{subfigure}{0.49\columnwidth}
        \centering
        \includegraphics[width=\columnwidth]{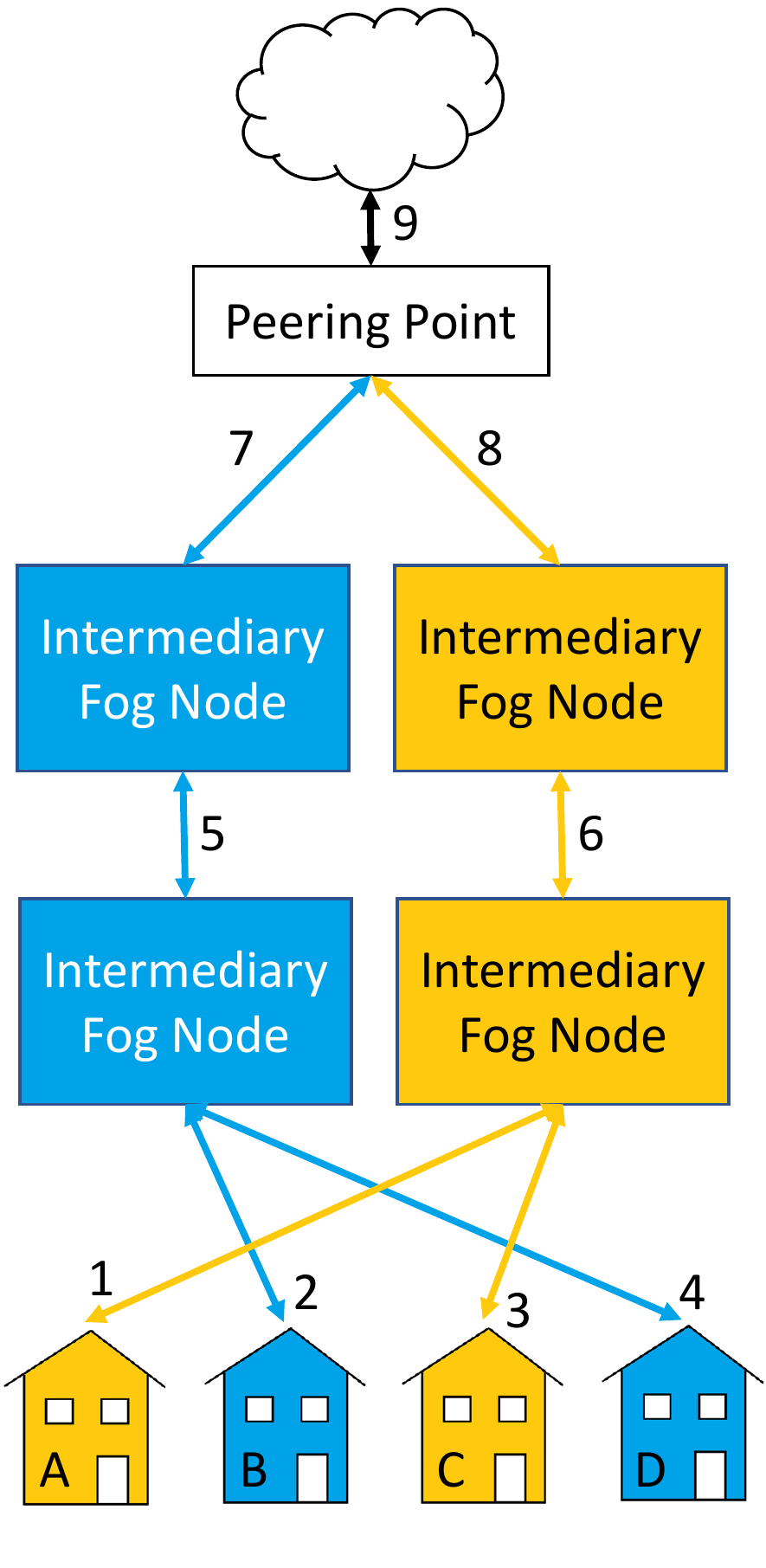}
        \caption{Message routing without grassroots peering.}
        \label{fig:routing-sota}
    \end{subfigure}
    \hfill
    \begin{subfigure}{0.49\columnwidth}
        \centering
        \includegraphics[width=\columnwidth]{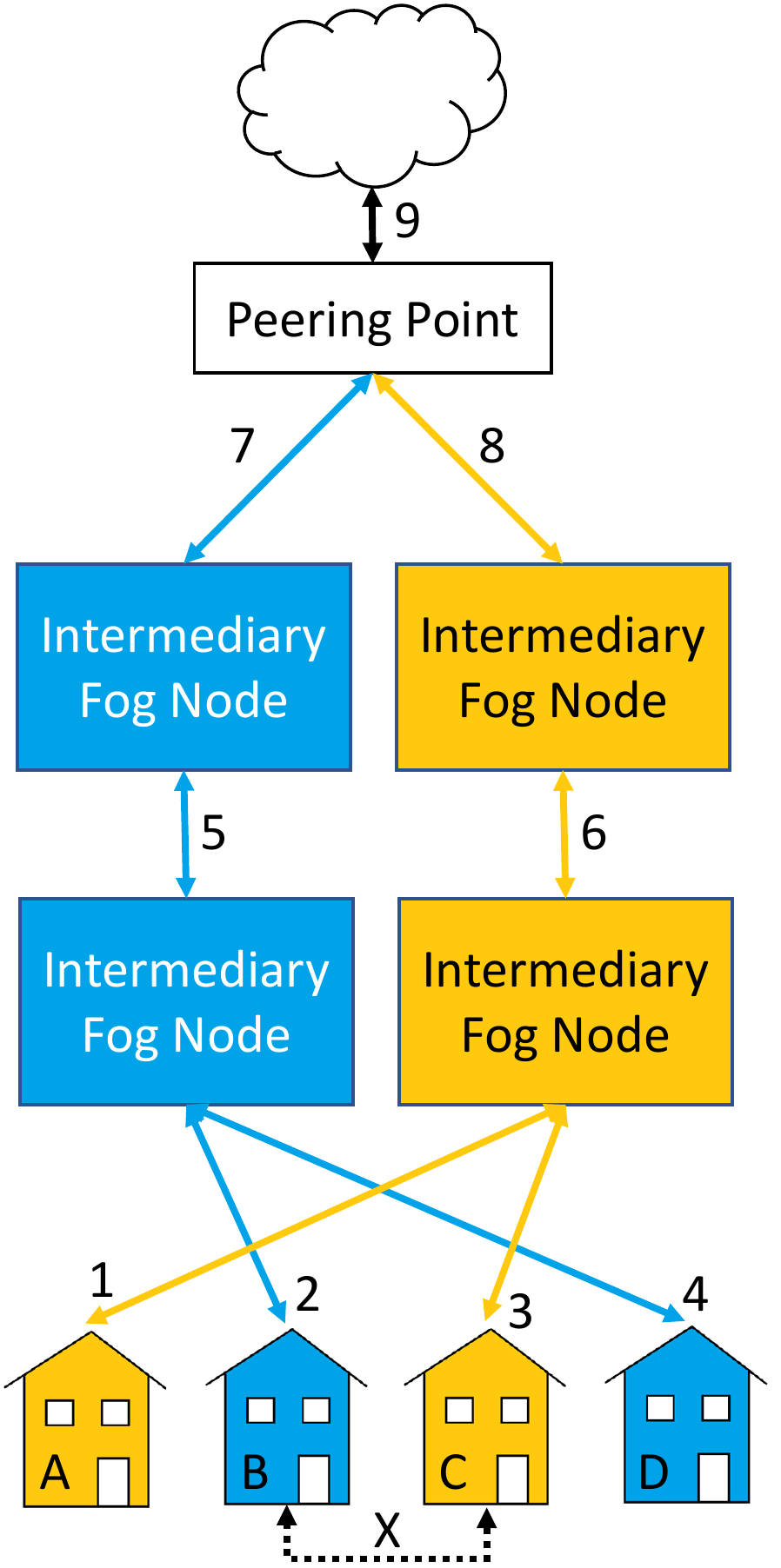}
        \caption{Message routing with grassroots peering.}
        \label{fig:routing-grassroots}
    \end{subfigure}
        \caption{Grassroots peering can provide shortcuts for parts of the cross-provider edge-to-edge traffic.}
    \label{fig:routing}
\end{figure}

Such use cases will often benefit strongly from direct edge-to-edge communication, e.g., in terms of low latency when a precipitation sensor is used to control windows or for privacy in the infant monitoring example.
Such direct edge-to-edge communication will usually be available in any setup resembling Figure~\ref{fig:overview-assumed}.
In a real setup, houses A and C in Figure~\ref{fig:routing-sota} can also have direct data sharing (in the abstract visualization simply the path (1-3)).
For houses A and B, however, data will follow the path (1-6-8-7-5-2) which is very obviously much longer and will usually result in higher latency.
Even if latency is not an issue, however, such a routing is inefficient: It is similar to shipping a parcel from one Berlin address to another via Beijing, Washington D.C. and Paris.

We argue that research should stop assuming low-latency edge-to-edge communication between all geographically close edge nodes and should instead focus more on the implications of multi-provider fog infrastructure.

\section{A Grassroots Approach for Cross-Provider Peering\label{sec:grassroots}}

In the long term, the ISPs will have to resolve this problem (and will if there is sufficient demand and, thus, a market).
The recent work on collaboration models and optimal resource allocation for federated edge resources~\cite{cao2020edgefederation,xavier2021edgefederation} highlights some of the challenges associated with solving the problem at the institutional level.  
In the short term, we propose that end users themselves create local peering points, thus, essentially following a grassroots approach for cross-provider peering.
In this section, we start by giving a brief overview of the proposed approach and of the enabling technology that already exists today before discussing the implications of our grassroots approach on the use case from Section~\ref{sec:problem}.

\subsection{Approach \& Enabling Technology}

The basic idea of grassroots peering is that edge operators in close proximity who, however, use different ISPs for Internet access, create an additional direct connection between their edge nodes as shown in Figure~\ref{fig:routing-grassroots}.

The main challenge is to establish communication links between pairs of edge devices while they are not initially aware of addressing or security information for the peers.
For example, in many residential settings, Wi-Fi access points can be a natural place for IoT data aggregation and exchange. 
While multiple neighboring access points can be in a communication range capable of supporting high-bitrate data exchange, information for bootstrapping such peering flows is typically missing.
One possible solution for exchanging the needed addressing and security information is for example the use of Wi-Fi beacon-stuffing~\cite{zehl2016lows}.
Alternatively, secondary communication interfaces with different radio technologies like ZigBee, LoRa, BLE can also be used as vectors.
Once this information is made available to the peers, it can be used to bootstrap direct data exchange~\cite{zehl2016resfi}.
This could also have a lot of potential when integrated with wireless community networks such as Freifunk\footnote{\url{https://freifunk.net/}}, NYC Mesh\footnote{\url{https://www.nycmesh.net/}}, or guifi.net\footnote{\url{https://guifi.net/}}.
Recent research has demonstrated even the possibility to create similar direct communication links between heterogeneous technologies, e.g. between Wi-Fi and LTE~\cite{gawlowicz2020punched}.

On the network level, the interaction can leverage standard protocols, for example IPv6-based routing combined with light-weight VPN tunneling using the WireGuard protocol\footnote{https://www.wireguard.com/} and centralized community-run key-exchange and coordination servers, similar to current commercial offerings like Tailscale\footnote{https://tailscale.com/}.

Finally, at the transport / application level we propose the use of an overlay network based on pub/sub, e.g., MQTT which is common in IoT use cases~\cite{paper_hasenburg_broadcast_groups}:
Both sides of the peering link operate a pub/sub broker and only messages originating from one of the brokers are allowed through the peering link.
Beyond this, these two can also decide to forward messages from other brokers:
in Figure~\ref{fig:routing-grassroots}, for instance, node C might decide to also forward messages from node A, received via the intermediary fog node.

\subsection{Implications for the Use Case}
For our use case, this means that all neighbors can still use their respectively own ISPs for Internet access but can share their IoT sensor data via more direct edge-to-edge channels.
Depending on the concrete implementation, this means that for the example deployment in Figure~\ref{fig:routing}, all path segments (5-7-8-6) and (6-8-7-5) are replaced by the X link, i.e., the connection of nodes B to C becomes (X) instead of (2-5-7-8-6-3) and for A to B becomes (1-3-X) instead of (1-6-8-7-5-2), thus, significantly shortening the edge-to-edge connection in our use case.

\section{Open Research Directions\label{sec:open}}
This opens up a number of open research directions which we discuss in this section.

\textbf{Implementing Grassroots Peering at the Edge:}
While we have discussed a few options regarding an actual implementation of grassroots peering at the edge, especially using overlay networks, it is still unclear how to actually implement this in practice.
We see a need for further research in this area, also to explore different implementation options and their respective advantages and disadvantages depending on concrete use case needs.

\textbf{Security and Privacy:}
The acceptance of the proposed grassroots peering will also crucially depend on careful selection of suitable security and privacy mechanisms like device identity management, privacy-enhancing layered encryption, etc.
The experience from nascent IoT overlay solutions like Amazon Sidewalk can serve as a valuable starting point.
Sidewalk minimizes the amount of metadata leakage between the application and the crowd-sourced network layer through the use of nested-encryption and reduces privacy risks like device\slash user\slash activity association by using ephemeral rolling transmission IDs~\cite{amazonsidewalk}.

\textbf{ISP Peering near the Edge:}
In the long term, it would be desirable to have the ISPs solve the issue of data sharing near the edge.
Aside from technical questions, we see a need for research regarding economics and incentives for providers to do this, i.e., it is not clear yet if and how ISPs will benefit from such peering.
In this context, sustainability incentives directed towards infrastructure sharing~\cite{govuk2021sharing} might play important role, based on further research on the lifetime environmental impact of edge hardware~\cite{skidar2013environmental}.

\textbf{Implications of Multi-Provider Environments on Existing Research:}
As outlined in our introduction, existing research on fog systems has so far mostly disregarded the problem of having multiple providers offer independent edge infrastructure near the edge.
We argue that researchers should carefully revisit their existing systems and designs and explore to which degree both the original problem solved as well as the solution will change in a multi-provider edge environment.
For instance, the keygroups of FBase~\cite{paper_hasenburg_towards_fbase,techreport_hasenburg_2019} need a low number of replicas to assert that the costs of maintaining data consistency remain reasonable.
If there are multiple ISPs with individual edge nodes in an area, keygroups should only be stored on the edge if they are used by a very small number of users.
All other keygroups in an optimal deployment will probably end up near the peering points of the respective ISPs.
Depending on the network infrastructure of the ISPs and the respective use case, it may even suffice to have one replica per ISP -- no matter if near the edge or not.

\section{Conclusion}
Fog computing, i.e., the combination of cloud, edge, and all resources on the edge-to-cloud path, allows applications to address their low latency and privacy requirements while coping with bandwidth limitations~\cite{paper_bermbach_fog_vision}.
Existing research on fog systems, however, has so far used a highly abstracted view on fog infrastructure.

In this position paper, we identified the problem of having separate independent provider networks for edge-to-edge communication and discussed its implications for edge-to-edge data sharing use cases.
As a mitigation approach, until ISPs solve this problem, we proposed a grassroots-based approach in which individual edge owners/operators install their own edge-to-edge communication network.
Based on this, we discussed a number of open research challenges -- both directly related to implementing our grassroots peering approach as well as for fog systems research built on top of such networks.

\bibliographystyle{plain}
\bibliography{cites}

\end{document}